# Isotope Effect in the Translation-Invariant Bipolaron Theory of High-Temperature Superconductivity

V.D. Lakhno

Keldysh Institute of Applied Mathematics of Russian Academy of Sciences, 125047 Moscow, Russia;

lak@impb.ru

**Abstract:** It is shown that the translation-invariant bipolaron theory of superconductivity can explain the dependence of the isotope coefficient in high-temperature superconductors on the critical temperature of a superconducting transition: in the case of strong electron–phonon interaction, the isotope coefficient is low when doping is optimal and high when it is weak. It is demonstrated that in the case of London penetration depth, the absolute value of the isotope coefficient behaves in the opposite way. A conclusion of the great role of non-adiabaticity in the case of weak doping is made. The criteria for d-wave phonon input into the isotope effect is established.

## 1. Introduction

Understanding the high-temperature superconductivity (HTSC) in cuprate superconductors and other HTSC materials is at the heart of current research in condensed matter physics. The isotope coefficient plays the central role in superconductivity (SC). The occurrence of the isotope effect was decisive in revealing the phonon mechanism of SC in ordinary superconductors. A lack of this effect in optimally doped high-temperature superconductors was the reason for discarding the phonon mechanism in HTSC and, as a consequence, the Bardeen–Cooper–Schrieffer theory (BCS) [1]. In recent years, however, a significant number of new experimental facts has led researchers to return to the electron–phonon interaction (EPI) as a dominant mechanism for explaining the HTSC effect. At the same time, the direct use of the BCS and its various modifications cannot explain these experimental facts [2].

The reason is probably that the BCS, being based on EPI, considers this interaction to be weak, while in the case of HTSC, it turns out to be strong. Generalization of the BCS to the case of strong EPI (Eliashberg theory) has been unable to explain many important phenomena attending HTSC, such as the pseudogap state. To overcome these difficulties, the author has developed a translation-invariant (TI) bipolaron theory of HTSC where the role of Cooper pairs belongs to TI bipolarons [3–6].

The aim of this paper is to explain isotope effects observed in HTSC on the basis of the TI bipolaron theory.

## 2. Isotope Influence on $T_c$

The isotope influence on the transition temperature $T_c$ played a decisive role in revealing the electron–phonon mechanism of a superconducting state and substantiating the Bardeen–Cooper–Schrieffer theory [1] for conventional superconductors. In the BCS theory, the isotope coefficient α for $T_c$ is determined from the relation found experimentally for ordinary metals, such that

$$T_c M^{\alpha} = const, \quad (1)$$

where $M$ is the mass of an atom replaced by its isotope. It follows from (1) that

$$\alpha = -dlnT_c/dlnM. \quad (2)$$

In the BCS theory, the value of α is positive and close to $\alpha \cong 0.5$, which is in good agreement with the experiment in ordinary metals. The great value of the isotope coefficient observed in ordinary metals implies a dominant role of EPI in them and suggests the applicability of the BCS theory for their description.

On the contrary, in high-temperature superconducting ceramics (HTSC), the isotope coefficient α is, generally, very small ($\sim 10^{-2}$) in the region of their optimal doping, which suggests that EPI is negligible and that other mechanisms of SC should be evoked [2].

As it is known, the BCS theory developed for the case of weak EPI is inapplicable in the case of HTSC where EPI is strong. In this case, the method of use may be the translation-invariant bipolaron theory developed in [3–6] (the reasons for which the Eliashberg theory [7], which is used in the case of strong EPI, can be unsuitable to describe HTSC are discussed in [6]).

According to the TI bipolaron theory, the temperature $T_c$ of a SC transition is determined by the following equation [3–6]:

$$T_c(\omega_0) = \left(F_{3/2}(0)/F_{3/2}(\omega_0/T_c)\right)^{2/3} T_c(0),$$

$$T_c(0) = 3.31\hbar^2 n_{b_p}^{2/3}/M_e, \; M_e = 2m, \quad (3)$$

$$F_{3/2}(x) = \frac{2}{\sqrt{\pi}}\int_0^{\infty}\frac{t^{1/2}dt}{e^{t+x}-1},$$

where $n_{b_p}$ is the concentration of TI bipolarons, $\omega_0$ is the frequency of an optical phonon, $m$ is the mass of a band electron (hole), ħ = h/$2\pi$, and h is the Planck constant. In the translation invariant bipolaron theory of superconductivity, the role of a SC gap belongs to a phonon frequency for which the electron–phonon interaction is maximum. In [3–6], consideration was mainly given to the isotropic law of phonon dispersion. In this paper, we will use them to calculate the isotope coefficient in HTSC materials (Appendix A). However, it can be shown that in the case of d-wave symmetry of the phonon spectrum, the results of [3–6] are valid (Appendix B). Expressions (2) and (3) yield the expression for the isotope coefficient:

$$\alpha = \frac{0.5}{1+\Phi(y)}, \qquad y = \frac{T_c}{\omega_0} \quad (4)$$

where the expression for $\Phi(y)$ is presented in Appendix A. It follows from (A6) that $\Phi(y) > 0$, so that $\alpha(y)$ is positive and lies in the interval $0 < \alpha(y) < 0.5$, where $\alpha = 0.5$ is the BCS value for the isotope coefficient.

The graph of the function $\alpha(T_c/\omega_0)$ is given in Figure 1.

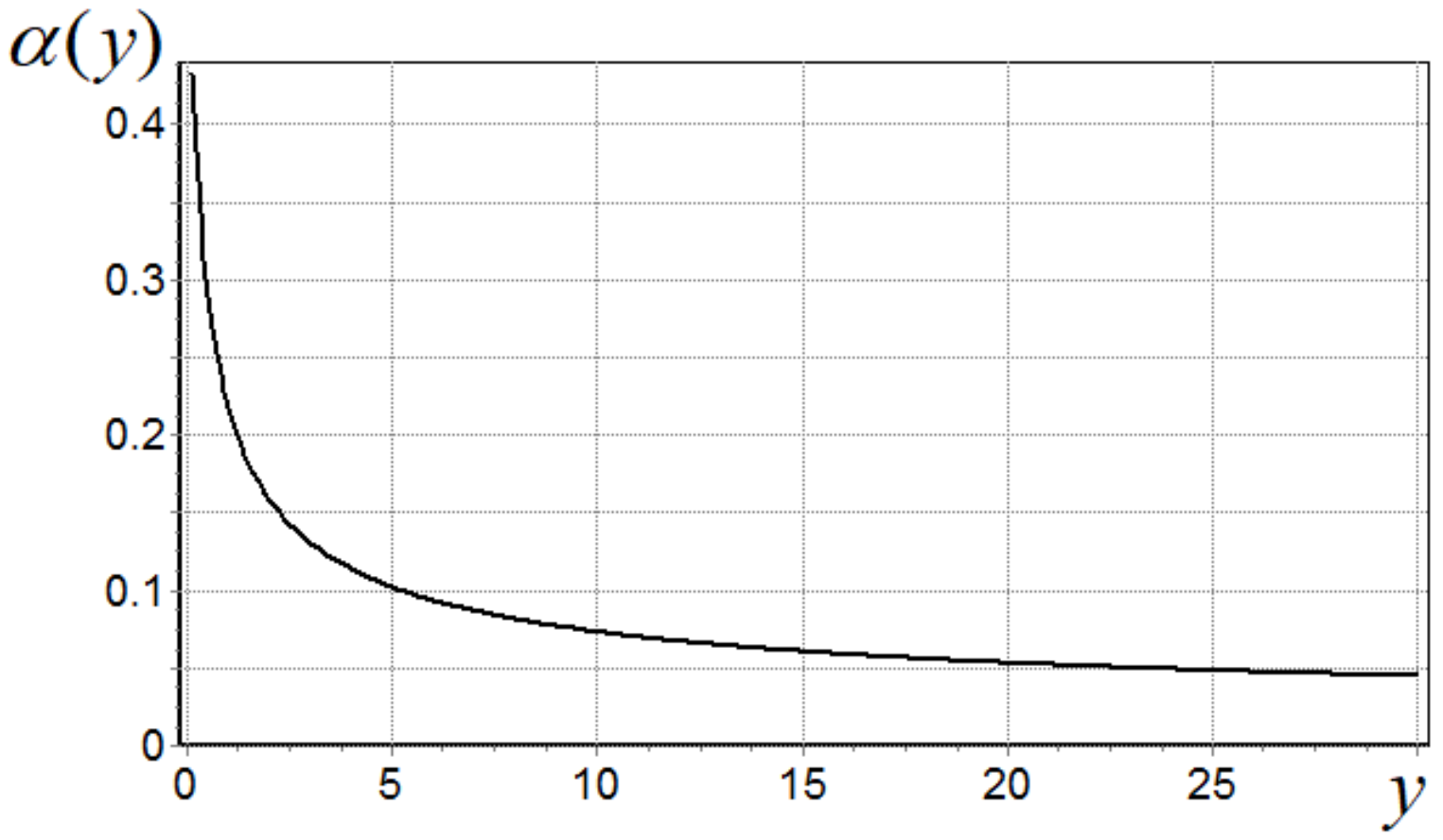

**Figure 1.** Dependence of the isotope coefficient α on the value of $y = T_c/\omega_0$.

The curve in Figure 1 shows that as the isotope coefficient decreases, the transition temperature increases. Figure 1 also suggests that in the case of $T_c/\omega_0 \gg 1$, which can correspond to optimally doped HTSC, the isotope coefficient will be small ($\alpha \to 0$ for $T_c/\omega_0 \to \infty$), in full agreement with the experiment [2,8–12].

In the opposite case: $T_c/\omega_0 \ll 1$, which corresponds to small doping; the isotope coefficient is maximum and equal to α = 0.5 as in the BCS theory.

Notice that according to (4), the isotope coefficients of different samples are the same for the same relations of $T_c/\omega_0$.

Figure 1 suggests that for typical values of α, the value of phonon frequency $\omega_0$ does not exceed $T_c$. For HTSC with $T_c$ = 100 K, this leads to $\omega_0$ being less than 8.6 meV.

## 3. Isotope Influence on London Penetration Depth

Quite a different picture is observed for the isotope coefficient for London penetration depth:

$$\beta = -\frac{M}{\lambda}\frac{d\lambda}{dM}, \qquad (5)$$

where $\lambda$ is London penetration depth:

$$\lambda = \left(\frac{M_e^* c^2}{16\pi e^2 n_0}\right)^{1/2}, \qquad (6)$$

$c$ is the velocity of light, $e$ is the electron charge, $M_e^*$ is the bipolaron mass, $n_0$ is the concentration of TI bipolarons in Bose condensate $n_0 = N_0/V$, and $N_0$ is the number of TI bipolarons in the condensate [3–6]:

$$\frac{N_0}{N} = 1 - \frac{\tilde{T}^{\frac{3}{2}}}{C_{bp}} F_{\frac{3}{2}}\left(\frac{\omega_0}{T}\right), \qquad (7)$$

$N$ is the total number of bipolarons, $C_{b_p} = \left(\frac{n^{2/3} 2\pi\hbar^2}{M_e \omega^*}\right)^{3/2}$, $\tilde{T} = T/\omega^*$, $\omega^*$ is the scale multiplier with energy dimension, and $n = N/V$.

As noted in [6], the mass of a TI bipolaron does not differ greatly from 2m, where m is the mass of a band electron which depends on $\omega_0$ only slightly. For this reason, we believe that the whole dependence on $\omega_0$ is determined by involving in (6) the concentration of TI bipolarons in the condensate $n_0$, related to $\omega_0$ by Formula (7). As a result, the equation below follows from (5)–(7) for the isotope coefficient $\beta$:

$$\beta = -\frac{1}{2}\frac{\tilde{\omega}_0 \tilde{T}^{1/2} N}{c_{b_p} N_0} Li_{1/2}\left(e^{-\omega_0/T}\right),$$
$$Li_{1/2}(Z) = \frac{1}{\sqrt{\pi}}\int_0^\infty \frac{1}{\sqrt{t}}\frac{dt}{Z^{-1}e^t - 1}. \quad (8)$$

It follows from (8) that in the limit of low temperatures when $\omega_0/T \gg 1, Li_{1/2}(e^{-\omega_0/T}) = e^{-\omega_0/T}$:

$$\beta = -\frac{1}{2}\frac{\tilde{\omega}_0 \tilde{T}^{1/2}}{c_{b_p}} e^{-\omega_0/T}, \quad (9)$$

that is, the isotope coefficient is exponentially small.

In the case of $\omega_0/T \ll 1, Li_{1/2}(e^{-\omega_0/T}) = \sqrt{\frac{\pi T}{\omega_0}}$, the isotope coefficient is equal to

$$\beta = -\frac{\sqrt{\pi}}{2}\frac{\tilde{\omega}_0^{1/2}\tilde{T}N}{N_0 c_{b_p}} = -\frac{\sqrt{\pi}}{2}\frac{\tilde{\omega}_0^{3/2} N}{N_0 c_{b_p}}\left(\frac{T}{\omega_0}\right). \quad (10)$$

It should be noted that as distinct from the coefficient α, which is positive, the $\beta$ for London penetration depth is negative, which is in agreement with the experiment. The fact that in the limit of low temperatures (9) the isotope coefficient $\beta$, caused by EPI, is negligible is consistent with the BCS [2]. Hence, in the limit of weak doping (when $T_c \to 0$), the isotope coefficient for London penetration depth $\beta$ as distinct from the isotope coefficient α for $T_c$ (which is high in this case) will be very small. The experiment, however, shows that the value of $\beta$ in the case of optimal doping (when $T_c$ is maximum) in the limit of low temperatures can be very large. For example, in optimally doped HTSC $YBa_2Cu_3O_7$ at $T = 0$ the value of $|\beta(0)| \approx 0.2$ [13]. Therefore, the main contribution in this case can be made by a non-adiabatic mechanism or non-phonon mechanisms [2,13]. An example of HTSC in which the contribution of non-adiabaticity and non-phonon mechanisms is probably small is provided by slightly overdoped $La_{2-x}Sr_xCu_{1-y}Zn_yO_4$, for which at $T = 0$, the isotope coefficient $\beta$ in accordance with the developed theory vanishes [14].

At high temperatures, on the contrary, the main contribution near $T_c$ can provide a phonon mechanism determined by (10) (according to [15], with the value α = 0.025 as observed in $YBa_2Cu_3O_{7-\delta}$, one obtains $\beta \sim -0.6$ for $T/T_c \sim 0.95$).

## 4. Discussion

The results obtained enable us to explain the peculiarities in the behavior of the isotope coefficient α observed for $T_c$ in high-temperature superconductors, in particular, its small value in the case of optimal doping and high value in the case of weak doping relying only on the electron–phonon interaction. From Figure 1, it can be observed that the critical temperature $T_c$ should be very large compared to phonon energy to explain the small values of isotope coefficient in the limit of optimal doping. This condition can be realized for soft optical phonon (Kohn anomaly [16–18]) for which TI bipolaron energy is the lowest (see Appendix B). The problem of the isotope coefficient $\beta$ for London penetration depth λ and its temperature dependence is more complicated. The electron–phonon interaction explains high values of $\beta$ for optimally doped HTSC materials only in the vicinity of the SC transition temperature $T_c$. In the case of low temperatures, the theory explains negligible values of $\beta(0)$ in HTSC $La_{2-x}Sr_xCu_{1-y}Zn_yO_4$. The experimental results for $La_{2-x}Sr_xCuO_4$ are often explained in terms of isotope dependence of carrier effective mass ($M_{bp}$ in Formula (6)):

$$\Delta M_{bp}/M_{bp} = 2\Delta\lambda/\lambda + \Delta n_0/n_0 \,.$$

It is stated that we can neglect the change in $n_0$ due to isotope effect, and all effects can be attribute to the change in effective mass $M_{bp}$ [19].

However, this statement is valid only in the case of applicability of the BCS theory in which the value $n_0$ coincides with the total number of normal phase electrons. As is shown in TI theory [3–6], $n_0$ represents only a small part of normal electrons. This is also confirmed by experiments of Božović et al. [20]. Thus, according to the results of Section 3, the isotope effect for London penetration depth can be explained by isotope dependence of $n_0$. However the theory does not explain the high values of $\beta(0)$ in other HTSC materials. In papers [2,13] this disagreement with the developed theory and the BCS is explained by the fact that in many HTSC materials at low temperatures, the main role belongs to non-adiabaticity effects, leading to high values of $\beta(0)$. With a simple model, such as that developed in this paper, one cannot hope to reproduce all the details (see, for example, [21–25]), and our goal was only the qualitative explanation of the most spectacular features. The theory cannot provide agreement for small doping where the experimental value for some samples is much larger than the BCS limit of $\alpha = 0.5$ and can have even negative values of $\alpha$. For an s-type order parameter, the isotope effect is always positive. The criteria for d-wave phonon input into the isotope effect when $\alpha$ can be negative is established in Appendix B.

It is important to mention that high-temperature superconductors are inherently inhomogeneous systems where translation invariance is absent. For the validity of BCS theory for HTSC, the characteristic size of inhomogeneity needs to be more than 1 micron due to the high value of the Cooper pair size. For translation-invariant theory, this size needs to be more than the bipolaron correlation length, which is about 1 nanometer. The measurement associated with surface and interfaces can reveal properties different from the bulk, the suppression of the d-wave order parameter and appearance of subdominant s-wave. Its inclusion in the theoretical model can greatly affect the obtained results.

**Appendix A. Derivation of Formula (4)**

Taking into account the relation $\omega_0 \sim M^{-1/2}$, we obtain from (2):

$$\alpha = \frac{\omega_0}{2T_c}\frac{dT_c}{d\omega_0}. \quad \text{A1)}$$

Let us introduce the notation $x = \omega_0/T_c$. Therefore,

$$\frac{dT_c}{d\omega_0} = \left(\frac{1}{T_c} - \frac{\omega_0}{T_c^2}\frac{dT_c}{d\omega_0}\right)\frac{dT_c}{dx}. \quad \text{A2)}$$

It follows from (A2) that

$$\frac{dT_c}{d\omega_0} = \frac{1}{T_c}\frac{1}{\left(1 + \frac{\omega_0}{T_c^2}\frac{dT_c}{dx}\right)}\frac{dT_c}{dx} \,. \quad \text{A3)}$$

Respectively, the expression (A1) for coefficient $\alpha$ will take the following form:

$$\alpha = \frac{\omega_0}{2T_c^2}\frac{1}{\left(1 + \frac{\omega_0}{T_c^2}\frac{dT_c}{dx}\right)}\frac{dT_c}{dx} \,. \quad \text{A4)}$$

Taking into account (3) for $dT_c/dx$, we obtain:

$$\frac{dT_c}{dx} = \frac{2}{3}\frac{F_{3/2}^{2/3}(0)}{F_{3/2}^{5/3}(x)}\ Li_{1/2}(e^{-x})\,, \qquad \text{A5)}$$

where $Li_{1/2}(x)$ is determined by (8). Setting $y = T_c/\omega_0$, we finally will obtain:

$$\alpha = \frac{0.5}{1+\Phi(y)},$$

$$\Phi(y) = 3y\int_0^\infty \frac{\sqrt{t}dt}{e^{(ty+1)/y-1}}\Big/\int_0^\infty \frac{dt}{\sqrt{t}(e^{(ty+1)/y-1})}. \qquad \text{A6)}$$

**Appendix B. The Criteria for D-Wave Phonon Input into Thermodynamic Properties of HTSC**

As is noted in [16], TI bipolarons form a charge density wave (CDW) with a wave vector equal to Fermi-surface nesting for which, due to Kohn anomaly, the phonon frequency vanishes. In HTSC materials, this is manifested in the availability of nodal directions in the dependence of the gap on the wave vector. In actual systems, the phonon frequency in the nodal direction, being small, does not turn to zero exactly. This is reflected in the dependence of the SC gap on the wave vector which in cuprate HTSC usually has the following form:

$$\omega_{CDW}(\vec{k}) = \Delta_0\left|cosk_x a - cosk_y a\right| + \omega_0, \qquad \text{(A7)}$$

where $\omega_{CDW}$, according to [16], is a phonon frequency corresponding to a nesting wave vector $Ƥ_{CDW}$ of a TI bipolaron. In the CDW theory, the first term in the right-hand side of (A7) corresponds to the contribution of the d-type wave, while the second term corresponds to the contribution of the s-type wave.

In calculating the thermodynamic properties of the TI bipolaron gas, the quantity $\omega_{CDW}$ is involved in the spectrum of TI bipolarons which, according to [16], has the following form:

$$\nu_k(Ƥ_{CDW}) = E_{bp}\Delta_{k,0} + \left(E_{bp} + \omega_{CDW}(\vec{k}) + \frac{k^2}{2M}\right)\left(1-\Delta_{k,0}\right), \qquad \text{(A8)}$$

where $\Delta_{k,0}= 1$ for $k = 0$ and $\Delta_{k,0} = 0$ for $k \neq 0$, $M$ = 2m, m is the mass of a band electron, and $E_{bp} = E_{bp}(Ƥ_{CDW})$ is the energy of the bipolaron ground state when its total momentum Ƥ is equal to $Ƥ_{CDW}$.

In calculating the statistical properties of an ideal TI bipolaron gas, which is a system of $N$, particles occurring in volume V we will proceed, as in [3–6], from the expression for N in the form:

$$N = N_0 + N',$$

$$N_0 = \frac{1}{e^{(E_{bp}-\mu)/T}-1},\ N' = \frac{V}{(2\pi\hbar)^3}\int d^3k\,\frac{1}{e^{(\nu_k-\mu)/T}-1}, \qquad \text{(A9)}$$

where $N_0$ is the number of TI bipolarons in the low one-particle state, and $N'$ is the number of TI bipolarons in higher states.

From (A7) and (A8), the condition where s-type CDW makes the main contribution into integral (A9) follows. Taking into account that the main contribution into the integral is made by $k \approx \sqrt{2MT}$ and $ka \ll 1$, from (A7) and (A8), we obtain the following: $\Delta_0\left|cosk_x a - cosk_y a\right|/T \cong \Delta_0 Ma^2/\hbar^2 \ll 1$. Hence, for $\omega_0 \gtrsim \Delta_0 Ma^2 T/\hbar^2$, the main contribution to the integral is made by the s-type SDW. In this case, the thermodynamic relations obtained in [3–6] remain unchanged. Thus, for example, in YBCO, the quantity $\frac{\omega_0}{\Delta_0} \approx 0.15$ [17,18], and the condition of applicability of s-approximation is fulfilled with great accuracy.